\documentstyle[aps,preprint,epsf]{revtex}
\newcommand{\lsim}{\stackrel{\scriptstyle <}{\phantom{}_{\sim}}}
\newcommand{\gsim}{\stackrel{\scriptstyle >}{\phantom{}_{\sim}}}

\begin{document}
\def\huge{\LARGE}
\title{\vspace*{-15mm}
Charge Screening at First Order Phase Transitions
}


\author{D. N. Voskresensky $^{1,2,3}$, M. Yasuhira
$^{1,4}$ and T. Tatsumi $^{4}$}

\maketitle

\noindent
$^{1}${\it Yukawa Institute for Theoretical Physics,  Kyoto 606-8502, Japan}
\\
$^{2}${\it
Gesellschaft f\"ur Schwerionenforschung (GSI),
Planck Str 1, D-64291 Darmstadt, Germany}
\\
$^{3}${\it
Moscow Institute for Physics and Engineering,
Kashirskoe sh. 31, Moscow 115409, Russia}
\\
$^{4}${\it Department of Physics, Kyoto University, Kyoto 606-8502, Japan}

\begin{abstract}
Possibility of structured mixed phases
at first order  phase
transitions  is examined
with taking into account of
charge screening and surface effects. Hadron-quark phase transition
in dense neutron star interior is considered, as
concrete example.
\end{abstract}

Ref. \cite{G92} suggested presence of a wide region
of mixed phase at any first order phase transitions in multi-component
systems of charged particles. Existence of structured
mixed phase in dense neutron star interiors would have important
consequences for equation of state, also
affecting neutrino emissivities \cite{RBP00},
glitch phenomena and $r$ modes, cf. \cite{BGP00,G01}.
However inhomogeneity effects of the field
profiles were  disregarded in these treatments.
On the other hand, ref. \cite{HPS93} demonstrated that
for the appearance of
the structured mixed phase the
Coulomb plus surface energy per droplet of
the new phase should have a minimum,
as function of the droplet radius.
Corrections  to
the Coulomb solutions due to screening effects were disregarded.
In spite of the question has been rised long ago, cf.
\cite{G92,RBP00,BGP00,G01,HPS93,NR00},
up to now there is no consistent treatment of these
effects.
Therefore, further study of the screening and surface
effects seems to be of prime importance.

Consider the structured mixed phase consisting of two phases I and II. We
suppose the lattice of droplets (phase I)
each placed in the Wigner-Seitz cell (the exterior of
droplets is phase II).
Each droplet in the cell
occupies the
domain $D^{\rm I}$ of volume $v^{\rm I}$
separated by a sharp boundary $\partial D$
from matter in
phase II ( region $D^{\rm II}$ of volume $v^{\rm II}$).
We exploit thermodynamic potential (effective energy) per cell
composed of a density
functional \cite{par},
\begin{eqnarray}
\Omega=E[\rho ] - \mu_i^{\rm I}\int_{D^{\rm I}}
d\vec{r}\rho_i^{\rm I}- \mu_i^{\rm II}\int_{D^{\rm II}}d\vec{r}
\rho_i^{\rm II} ,
\label{omeg}
\end{eqnarray}
$E[\rho ]$
is the energy of the cell, $\rho =\{\rho_i^{\rm I},  \rho_i^{\rm II}\}$
are densities of different particle species, $i=1,...,N^{\rm I}$ in phase I
and $i=1,...,N^{\rm II}$ in phase II, $N^{\rm I}$,  $N^{\rm II}$
are  total number of
particle species per cell in phases I and II.
Summation over the repeated Latin indices is implied.
Chemical potentials $\mu_i^{\rm I}$, $\mu_i^{\rm II}$
are
constants, if, as we assume, each phase is in the ground state.
We also assume that matter in phase I or II is in chemical equilibrium by
means of the weak and strong interactions.
Equations  of motion, $\frac{\delta\Omega}{\delta\rho_i^\alpha}=0$, render
\begin{eqnarray}\label{omeg-chem}
\mu_i^\alpha  = \frac{\delta E [\rho ]}{\delta \rho_i^{\alpha}}, \,\,\,\alpha =
\{{\rm I}, {\rm II}\} .
\end{eqnarray}

The energy of the cell consists of four contributions:
\begin{equation}\label{enden1}
E
[\rho ] =\int_{D^{\rm I}} d\vec{r} \epsilon^{\rm I}_{\rm kin+str}
[\rho_i^{\rm I}] +
\int_{D^{\rm II}} d\vec{r} \epsilon^{\rm II}_{\rm kin+str}
[\rho_i^{\rm II}]+\int_{\partial D} dS\epsilon_S [\rho ]+E_V.
\end{equation}
The first two contributions are the sums of the kinetic
and strong-interaction energies  and
$\epsilon_S [\rho ]$
is the surface energy density, which depends on all the particle
densities at the boundary $\partial D$. One may approximate it, as
we shall do, in terms
of surface tension $\sigma$.
\footnote{In reality there is no {\it sharp boundary} between phases
and
all the densities
and constant chemical potentials
are defined in the whole space
(c.f. models of atomic nuclei).
Then  particle
densities, $\rho^\alpha_i$, are
changed continuously.
The surface energy is given by integration of
$\epsilon^\alpha$ over a narrow region around  surface,
where $\rho_i^\alpha$ change sharply.
Our formulation is also applicable to this case.}
$E_V$ is the Coulomb
interaction energy,
\begin{equation}\label{enden2}
E_V=\frac{1}{2}\int d\vec{r}\,d\vec{r}^{\,\prime}
\frac{Q_i \rho_i (\vec{r}) Q_j
\rho_j (\vec{r}^{\,\prime}) }{\mid
\vec{r}-\vec{r}^{\,\prime}\mid}
,
\end{equation}
with $Q_i$ being the particle
charge
($Q =-e <0$ for the electron).

Then equations of motion (\ref{omeg-chem}) can be  re-written as
\begin{equation}\label{eom}
\mu_i^\alpha =\frac{\partial\epsilon_{\rm kin+str}^\alpha}
{\partial\rho_i^\alpha}-
N^{{\rm{ch}},\alpha }_i (V^\alpha -V^0 )  ,~~~
N^{{\rm{ch}},\alpha}_i =Q_i^\alpha /e,
\end{equation}
where $V^\alpha (\vec r )$ is the electric potential generated
by the particle distributions,
which
can be shifted by an arbitrary constant ($V^0$) due to the
gauge transformation, $V\rightarrow V -V^0$,
\begin{eqnarray}\label{other1}
V(\vec r )
=- \int d\vec{r}^{\,\prime}\,
\frac{e Q_i \rho_i (\vec{r}^{\,\prime}) }{\mid
\vec{r}-\vec{r}^{\,\,\prime}\mid }+V^0
\equiv\left\{
\begin{array}{ll}
V^{\rm I}(\vec{r}), & \vec{r}\in D^{\rm I}\\
V^{\rm II}(\vec{r}), & \vec{r}\in D^{\rm II}
\end{array}
\right.
\end{eqnarray}
Formally varying
eq.~(\ref{eom}) with respect to $V^\alpha$ or $\mu_i^\alpha$
we have the matrix form relation,
\begin{eqnarray}\label{matrix}
A_{ij}^\alpha\frac{\partial\rho_{j}^\alpha}{\partial V^\alpha}= N^{{\rm ch},\alpha}_i ,\,\,\,\,
A_{ij}^\alpha B_{jk}^\alpha =\delta_{ik},
\end{eqnarray}
where  matrices $A$ and $B$ are defined as
\begin{eqnarray}
A_{ij}^\alpha \equiv\frac{\delta^2 E_{\rm kin+str}^\alpha}{\delta\rho_i^\alpha\delta\rho_j^\alpha},\,\,\,\,
B_{ij}^\alpha\equiv\frac{\partial\rho_i^\alpha}{\partial\mu_j^\alpha}.
\label{matrix1}
\end{eqnarray}
Eqs.~(\ref{matrix}), (\ref{matrix1})  reproduce
gauge-invariance relation,
\begin{equation}
\frac{\partial\rho_i^\alpha}{\partial V^\alpha}=N^{{\rm ch},\alpha}_j
\frac{\partial\rho_j^\alpha}{\partial\mu_i^\alpha},
\label{gauge}
\end{equation}
clearly showing that constant-shift of the chemical
potential is
compensated by gauge transformation of $V^\alpha$. Hence chemical potential
$\mu_i^\alpha$
acquires physical meaning only after fixing of the gauge of $V^\alpha$.

Applying Laplacian ($\Delta$) to the l.h.s. of eq.~(\ref{other1}) we recover
the Poisson equation ($\vec{r}\in D^\alpha$),


\begin{eqnarray}\label{qV}
\Delta V^\alpha =4\pi e^2\rho^{{\rm ch},\alpha}\equiv 4\pi e Q_i^\alpha
\rho_i^\alpha
. \,\,\,
\end{eqnarray}
The charge density $\rho^{{\rm ch},\alpha}$  as a function of
$V^\alpha$ is determined
by equations of motion (\ref{eom}), so that
eq.~(\ref{qV}) is a nonlinear differential equation for $V^\alpha$.
The boundary conditions are
\begin{equation}
V^{\rm I}=V^{\rm II},~~~\nabla V^{\rm I}=\nabla V^{\rm II},~~~
{\vec r}\in \partial D\, ,
\end{equation}
where we neglected a small contribution of  surface charge
accumulated at the interface of the phases. We also impose
condition $\nabla V^{\rm II}=0$ at the boundary of the Wigner-Seitz cell,
which implies that each cell must be charge-neutral.
Once eqs.~(\ref{qV}) are solved giving $V^\alpha$ and the
potentials are matched
at the boundary,
we have density
distributions of particles in the domain $D^\alpha$.

Note that there are two
conservation laws relevant in neutron star matter: baryon number and charge
conservation. These quantities are well defined over the whole space,
not
restricted to each domain. Accordingly the baryon number
and charge
chemical potentials ($\mu_B$ and $\mu_Q$) , being linear combinations of
$\mu_i^\alpha$,
become constants over the whole space,
\begin{equation}\label{gibbs1}
\mu^{\rm I}_B=\mu^{\rm II}_B\equiv \mu_B,~~~\mu^{\rm I}_Q=\mu^{\rm II}_Q
\equiv \mu_Q .
\end{equation}
 This fact requires two conditions for $\mu_i^\alpha$
at the boundary $\partial D$,
which prescribe the conversion manner
of particle species of two phases at the interface. In other words, charge
and baryon number densities should be continuous across the boundary due to
eq. (\ref{eom}).
\footnote{Each particle density is not necessarily continuous across the
boundary, since it is only defined in each phase, while densities of
leptons are well defined over the whole space. When particles of
the same species $i$ are allocated in both domains and the conversion of
particle species becomes trivial, we must further impose the relations,
$\mu_i^{\rm I}=\mu_i^{\rm II}$, and $\rho_i^{\rm I}=\rho_i^{\rm II}$
at the boundary.}
In particular, electron chemical potential is equal to the charge chemical
potential ($\mu_Q=\mu_e^\alpha$) and its number density is related as
\begin{equation}
\rho_e^\alpha=\frac{(\mu_Q-V^\alpha+V^0)^3}{3\pi^2},
\end{equation}
from eq.~(\ref{eom}). Note that this is a gauge-invariant quantity.

Once eq.~(\ref{eom}) is satisfied, pressure becomes constant in each domain,
\begin{eqnarray}\label{pressure}
-P^\alpha v^\alpha&=&\int_{D^\alpha} d{\vec r}
\left\{\epsilon^\alpha_{\rm kin+str}
[\rho^\alpha_i]-\frac{1}{2}N_i^\alpha\rho_i^\alpha V^\alpha
-\mu_i^\alpha\rho^\alpha_i\right\}.
\end{eqnarray}
Hence, the
extremum condition for $\Omega$ with respect to a modification
of the boundary of arbitrary shape (under the total volume of the
Wigner-Seitz cell being fixed) reads
\begin{equation}\label{gibbs2}
P^{\rm I}=P^{\rm II}+\sigma\frac{d S}{dv^{\rm I}},
\end{equation}
$S$ is the area of the boundary $\partial D$ and
$\sigma$ is surface tension. The boundary of the cell does not contribute since
all the densities are continuous quantities at this point.
Eq. (\ref{gibbs2}) is the pressure equilibrium condition between two phases.
\footnote{
As
noted in footnote 1, in realistic problem
with continuous density distributions, i.e. in absence of sharp boundary,
the contribution of the surface energy is
absorbed into $P^\alpha$.
Hence $P^{\rm I}=P^{\rm II}$ in such a more detailed treatment.}
{\it Thus we  satisfy Gibbs conditions (\ref{gibbs1}),
(\ref{gibbs2}) in our formalism.}

The Debye screening parameter is determined by
the Poisson
equation if one expands
the charge density in
$\delta V^\alpha =V^\alpha -V^{\alpha}_{\rm r}$ around a reference value
$V^{\alpha}_{\rm r}$.
Then eq.~(\ref{qV}) renders
\begin{eqnarray}\label{Pois-lin}
&&\Delta \delta V^\alpha
=  4\pi e^2
\rho^{{\rm ch},\alpha} (V^{\alpha}=V^{\alpha}_{\rm r} )+
(\kappa^{\alpha} (V^{\alpha}=V^{\alpha}_{\rm r} ))^2 \delta V^{\alpha}
\\
&&+
2\pi e^2\left[ \frac{\partial^2\rho^{{\rm ch},\alpha}}{(\partial V^{\alpha})^2}\right]_
{V^{\alpha}=V^{\alpha}_{\rm r}}(\delta V^{\alpha})^2 +...
, \nonumber
\end{eqnarray}
\begin{equation}\label{debye}
(\kappa^{\alpha}(V^{\alpha}=V^{\alpha}_{\rm r} ))^2 =
4\pi e^2\left[ \frac{\partial\rho^{{\rm ch},\alpha}}{\partial V}\right]_
{V^{\alpha}=V^{\alpha}_{\rm r}}
=4\left.\pi
Q_i^{\alpha}Q_j^{\alpha}\frac{\partial\rho_j^{\alpha}}{\partial\mu_i^\alpha}
\right|_{V^{\alpha}=V^{\alpha}_{\rm r}},
\end{equation}
where we used eq.~(\ref{gauge}).
Then we calculate contribution to the thermodynamic potential
(effective energy) of the cell up to $O((\delta V^{\alpha})^{2})$. A
proper electric field energy of the cell is
\begin{eqnarray}\label{eV}
E_V
=\int_{D^{\rm I}} d{\vec r}\epsilon_{V}^{\rm I}
+\int_{D^{\rm II}}d{\vec r}\epsilon_{V}^{\rm II}
=\int_{D^{\rm I}}\frac{(\nabla
V^{\rm I})^2}{8\pi e^2}d{\vec r} + \int_{D^{\rm II}}\frac{(\nabla
V^{\rm II})^2}{8\pi e^2}d{\vec r}  ,
\end{eqnarray}
that  in the case of  unscreened
distributions is usually called the Coulomb energy.
Besides the terms given by (\ref{eV}), there are another contributions
arising from effects associated with inhomogeneity of the
electric potential profile, due to implicit dependence
of partial contributions to the particle densities on $V^{\rm I(II)}$.
We will call them correlation terms.
Then taking $\rho_{i}^{\alpha}$ as function of $V^{\alpha}$
we expand
$\epsilon_{\rm kin+str}^{\alpha}$ in $\delta V^{\alpha}$:
\begin{eqnarray}\label{exp3}
&&
\epsilon_{\rm kin+str}^{\alpha}[\rho ]=
\epsilon_{\rm kin+str}^{\alpha}(\rho_i^{\alpha}(V^{\alpha}
))+\left[ \left(\mu_i^\alpha + N^{{\rm ch},{\alpha}}_i (V^{\alpha} -V^0
)\right)
\frac{\partial\rho_i^{\alpha}}{\partial V^\alpha }\right]_{V^{\alpha}=
V^{\alpha}_{\rm r}}\delta V^{\alpha}
\nonumber
\\
&&
+\frac{1}{2}\left[ \frac{(\kappa^{\alpha})^2}{4\pi e^2}
+\left(\mu_i^\alpha + N^{{\rm ch},{\alpha}}_i (V^{\alpha} -V^0
)\right)\frac{\partial^2\rho_i^{\alpha}}{(\partial V^{\alpha})^2}
\right]_{V^{\alpha}=V^{\alpha}_{\rm r}}(\delta V^{\alpha})^2 + ...
\end{eqnarray}
We used eqs.~(\ref{omeg-chem}), (\ref{other1}), (\ref{matrix}) and
(\ref{matrix1}) in this derivation. The term in (\ref{exp3}) $\propto (V^{\alpha} -V^0 )(\delta V^{\alpha})^2 $
is actually of the higher order of smallness than other terms if $V^{\alpha} \sim V_0 \sim V^{\alpha}_{\rm
r}$.
Using expansion
\begin{eqnarray}
-\mu_i^\alpha \rho_i^{\alpha}
=-[\mu_i^\alpha \rho_i^{\alpha} ]_{V^{\alpha}_{\rm r}}
- \left[\mu_i^\alpha\frac{\partial \rho_i^{\alpha} }{\partial V^{\alpha}}
\right]_{V^{\alpha}_{\rm r}}
\delta V^{\alpha}
-\frac{1}{2} \left[\mu_i^\alpha\frac{\partial^2 \rho_i^{\alpha} }
{(\partial V^{\alpha})^2}\right]_{V^{\alpha}_{\rm r}}
(\delta V^{\alpha})^2 +...
\end{eqnarray}
we obtain the corresponding
correlation contribution to the thermodynamic potential $\Omega_{\rm cor}
=\int_{D^{\rm I}} d{\vec r}\omega_{\rm cor}^{\rm I}
+\int_{D^{\rm II}} d{\vec r}\omega_{\rm cor}^{\rm II}$:
\begin{eqnarray}\label{om-cor0}
&&\omega_{\rm cor}^{\alpha}=
\epsilon_{\rm kin+str}^{\alpha}(\rho_i^{\alpha}(V^{\alpha}_{\rm
r}))-\mu_i^\alpha \rho_i^{\alpha} (V^{\alpha}_{\rm r})-\rho^{{\rm ch},\alpha} (V^{\alpha}_{\rm r} )
(V^{\alpha}_{\rm r} -V^0 )
\nonumber \\
&&+\frac{(V^{\alpha}_{\rm r} -V^0 )\Delta V^{\alpha}}{4\pi e^2}
+\frac{
(\kappa^{\alpha}(V^{\alpha}_{\rm r }) )^2
(\delta V^{\alpha})^2 }{8\pi e^2 }+...
\end{eqnarray}
where we also used eqs. (\ref{Pois-lin}) and (\ref{debye}).
In general $V^{\rm I}_{\rm r }\neq V^{\rm II}_{\rm r }$ and they may depend
on the droplet size.  Their choice should provide appropriate convergence of the
above
expansion in $\delta V$. Taking $V^{\rm I}_{\rm r }= V^{\rm II}_{\rm r }=V_{\rm r } = const$
we find
\begin{eqnarray}\label{om-cor}
\omega_{\rm cor}^{\alpha}=\frac{
(\kappa^{\alpha}(V_{\rm r }) )^2
( V^{\alpha} - V_{\rm r })^2 }{8\pi e^2 } + const ,
\end{eqnarray}
and one may count the potential from the corresponding constant value.

In the following we consider the hadron-quark phase transition, as an example.
We suppose the lattice of spherical droplets of the radius $R$ placed in the
Wigner-Seitz cell of the radius $R_{\rm W}$.
We also assume the quark matter inside the droplet, as phase I,
and the hadronic matter outside in the cell, as phase II,
both divided by a sharp
boundary $r=R$. The quark matter consists of $u,d,s$ quarks and electrons
and the kinetic plus strong interaction energy density is given by
\begin{eqnarray}\label{endenq1}
\epsilon_{\rm kin+str}^{\rm I} \simeq
\frac{3\pi^{2/3}}{4}\left(
1+\frac{2\alpha_c }{3\pi} \right)
\left[ \rho_u^{4/3}+\rho_d^{4/3}+\rho_s^{4/3}
+\frac{m_s^2\rho_s^{2/3}}{\pi^{4/3}}
\right]
+B+\frac{(3\pi^2 \rho_e)^{4/3}}{4\pi^2}, \nonumber
\end{eqnarray}
where $B$ is the bag constant, $\alpha_c $
is the QCD coupling constant, $m_s $
is the mass of strange quark.
Last term is the kinetic energy of electrons.

The hadronic matter consists of protons,
neutrons and electrons and the kinetic plus strong
energy density is given by
\begin{equation}\label{eVh11}
\epsilon_{\rm kin+str}^{\rm II}\simeq \epsilon_n^{\rm kin}[\rho_n ]+ \epsilon_p^{\rm kin}[\rho_p ]+
\epsilon_{\rm pot}[\rho_n,\rho_p]+\frac{(3\pi^2 \rho_e)^{4/3}}{4\pi^2},
\end{equation}
where  $\epsilon_i^{\rm kin}[\rho_i ], i=n,p$ are standard relativistic
kinetic energies of
nucleons, while $\epsilon_{\rm pot}$ is the potential energy contribution
we take
here in the form
\begin{eqnarray}\label{eVh11p}
\epsilon_{\rm pot}[\rho_n,\rho_p]&=&S_0 \frac{(\rho_n - \rho_p )^2}{\rho_0}+
(\rho_n +\rho_p ) \epsilon_{\rm bind} +\frac{K_0 (\rho_n +\rho_p ) }{18}
\left( \frac{\rho_n +\rho_p }{\rho_0}-1 \right)^2 \nonumber\\
 &+& C_{\rm sat} (\rho_n +\rho_p ) \left( \frac{\rho_n +\rho_p }
{\rho_0} -1 \right)  ,
\end{eqnarray}
$\rho_0$ is the nuclear density ($\rho_0\simeq 0.16$fm$^{-3}$) and
constants $\epsilon_{\rm bind}$, $K_0$, $C_{\rm sat}$ are determined to
satisfy the nuclear saturation properties.

We use chemical equilibrium conditions for the reactions
$u+e \leftrightarrow s$,
$d\leftrightarrow s$,
and $n\leftrightarrow p+e$ in each phase,
\begin{eqnarray}\label{q-h}
\mu_u -\mu_s +\mu_e =0, \,\,\,\, \mu_d =\mu_s, \,\,\,\,
\mu_n =\mu_p +\mu_e ,
\end{eqnarray}
and the conversion relation at the boundary,
\begin{eqnarray}\label{q-h1}
\mu_B\equiv \mu_n =2\mu_d +\mu_u ,
\end{eqnarray}
which yield relations between quark and nucleon chemical potentials.
\footnote{Other conversion
relation $\mu_p =2\mu_u +\mu_d$ is then automatically satisfied.}
Using these
conditions
we obtain, cf. \cite{HPS93},
\begin{eqnarray}\label{qrhoch}
\rho^{\rm ch ,I}\simeq \left( 1-\frac{2\alpha_c}{\pi}
\right) \left[ \frac{2\mu_B^2 V^{\rm I}}{9\pi^2 }
\left(1+O\left(\left(\frac{V^{\rm I}
}{\mu_B}
\right)^2\right)
\right)+
\frac{\mu_B  m_s^2 }{6\pi^2 } \right] ,
\end{eqnarray}
where the electron contribution is omitted as small ($\sim
(V^{\rm I})^3$) and we fixed the gauge by taking $V^0 =-\mu_Q \equiv -\mu_e$.


Poisson equation (\ref{Pois-lin}) with $\rho^{\rm ch ,I}$ from (\ref{qrhoch})
describing electric potential of
the quark droplet  can be solved analytically.
For $r<R$ with the boundary condition
$|V^{\rm I}(r\rightarrow 0)|<\infty$, we
find
\begin{equation}\label{1-sol}
V^{\rm I}=\frac{V_{0}^{\rm I}}{\kappa^{\rm I} r}{\rm sh}(\kappa^{\rm I} r)+
U_{0}^{\rm I},
\end{equation}
with an arbitrary constant $V_{0}^{\rm I}$.
For  the Debye parameter $\kappa^{\rm I}$
and for the constant
$U_{0}^{\rm I}$ we obtain:
\begin{eqnarray}\label{qscrl}
(\kappa^{\rm I})^2 =\frac{8e^2 \mu_{B}^2}{9\pi}\left( 1-
\frac{2\alpha_c}{\pi}
\right),\,\,\,\, U_{0}^{\rm I} \simeq -\frac{3 m_s^2}{4\mu_B}.
\end{eqnarray}
Note that solution (\ref{1-sol}) is independent of the reference
value $V_{\rm r}^{\rm I}$ in this case, c.f. (\ref{debye}), since
$\rho^{\rm ch, I}$ in (\ref{qrhoch}) is  linear function of
$V^{\rm I}$ in approximation used.

For phase II, expanding the charge density $\rho^{\rm ch ,II}$
around a reference value, $\rho^{\rm ch ,II}\simeq\rho_p (V^{\rm
II}=V_{\rm r}^{\rm II} )+\delta \rho_p -\rho_e(V^{\rm II}=V_{\rm
r}^{\rm II})-\delta\rho_e$, and using eqs.~(\ref{matrix}),
(\ref{matrix1}), (\ref{debye}) and (\ref{eVh11}) we find up to
linear order
\begin{eqnarray}\label{co}
&&\delta \rho_p \simeq C_0^{-1}(V^{\rm II}(r)-V_{\rm r}^{\rm II}),
\,\,\,\, \delta\rho_e=\frac{(V_r^{\rm II})^2}{\pi^2} (V^{\rm
II}(r)-V_{\rm r}^{\rm II})\\ &&C_0=\frac{A_{22}}{|A|} \simeq
\frac{\pi^2}{p_{{\rm F}p}\sqrt{p_{{\rm
F}p}^2+m^2}}+\frac{4S_0}{\rho_0},~~~ p_{{\rm F}p}=(3\pi^2\rho_p
(V^{\rm II}=V_{\rm r}^{\rm II}))^{1/3}.\nonumber
\end{eqnarray}

For $r>R$, the Poisson equation with the boundary condition
$V^{\prime}|_{R_{\rm W}}=0$  yields
\begin{eqnarray}\label{psi}
V^{\rm II}= V_{0}^{\rm II}
\frac{R}{r} \mbox{ch}\left(\kappa^{\rm II}(r-R_{\rm W} )\right)
\left( 1-\delta\right)+U_0^{\rm II}
\end{eqnarray}
with an arbitrary constant $V_0^{\rm II}$, where the constant
$U_0^{\rm II}$ is given by
\begin{equation}
U_0^{\rm II}=-\frac{4\pi e^2\rho^{\rm ch, II}(V^{\rm II}=V_r^{\rm II})}
{(\kappa^{\rm II})^2}+V_r^{\rm II}.
\end{equation}
We will further drop numerically small
$\delta$ correction.
We take the reference value $V_{\rm r}^{\rm II}= V(r=R_{\rm W})$.
Since size of the Wigner-Seitz cell  $R_{\rm W}$ is substantially
larger than $R$, we have $V(r=R_{\rm W}) \simeq V^{\rm bulk}$,
where
$V^{\rm bulk}$ is constant bulk solution of the Poisson equation,
$(V^{\rm bulk})^3 \equiv -\mu_{e}^3=-3\pi^2
\rho_p (V^{\rm bulk}=-\mu_e )$,
that coincides with the local
charge-neutrality condition for the case of the spatially homogeneous
matter. Hence we find $U_0^{\rm II}\simeq V_{r}^{\rm II}\simeq -\mu_e$.
\footnote{Thus solutions (\ref{1-sol}) and (\ref{psi})
are proved to be
consistent with eq.~(\ref{other1}) for the gauge choice, $V^0=-\mu_e$.}
The charge screening in the external region is
determined by the Debye parameter
\begin{eqnarray}\label{hscrl}
(\kappa^{\rm II})^2 =\frac{4e^2 \mu_{e}^{2}}{\pi}
+\frac{4e^2 \pi}{C_0},
\end{eqnarray}
where second term is contribution of proton screening.
Taking $\rho_{B}^{\rm II}=1.5 \rho_0$,   $\mu_{e}
\simeq 170$~MeV, $\mu_{B}=\mu_n \simeq 1020$~MeV,
$\alpha_c \simeq 0.4$,
we roughly estimate typical Debye screening lengths as
$\lambda_{D}^{\rm I}\equiv 1/\kappa^{\rm I}\simeq 3.4/m_\pi$, and
$\lambda_{D}^{\rm II}\equiv 1/\kappa^{\rm II}
\simeq 4.2 /m_\pi$, whereas one would have
$\lambda_{D}^{\rm II}\simeq 8.5 /m_\pi$,
if the proton contribution to
the screening (\ref{hscrl})  was absent ($C_0^{-1}=0$).

Matching of the fields yields
\begin{eqnarray}\label{const1}
V_{0}^{\rm I} \simeq \frac{ \left(U_0^{\rm II}-U_0^{\rm I}
\right)\left[\mbox{ch}(\kappa^{\rm II}(R_{\rm W} -R))+\kappa^{\rm II}R\,
\mbox{sh}(\kappa^{\rm II}(R_{\rm W} -R))\right]}{\alpha_0 \mbox{sh}(\kappa^{\rm I}R)
\mbox{sh}(\kappa^{\rm II}(R_{\rm W} -R))+\mbox{ch}(\kappa^{\rm I}R)
\mbox{ch}(\kappa^{\rm II}(R_{\rm W} -R))
},
\end{eqnarray}
\begin{eqnarray}\label{const2}
V_{0}^{\rm II}\simeq -\frac{ \left(U_0^{\rm II}-U_0^{\rm I}
\right)\left[\mbox{ch}(\kappa^{\rm I}R)-
\mbox{sh}(\kappa^{\rm I}R)/(\kappa^{\rm I}R)\right]}{\alpha_0 \mbox{sh}(
\kappa^{\rm I}R)
\mbox{sh}(\kappa^{\rm II}(R_{\rm W} -R))+\mbox{ch}(\kappa^{\rm I}R)
\mbox{ch}(\kappa^{\rm II}(R_{\rm W} -R))
},
\end{eqnarray}
where we introduced  notation
$\alpha_0 = \kappa^{\rm II}/\kappa^{\rm I}$.

The charge in the sphere of current radius $r<R$ is given by
\begin{eqnarray}
Q(r) =V_{0}^{\rm I}~r \mbox{ch}(\kappa^{\rm I}r)\left(
1-\frac{\mbox{th}(\kappa^{\rm I}r)}{\kappa^{\rm I}r}\right)<0 ,
\end{eqnarray}
being, thereby, negative, since
$U_0^{\rm II}>U_0^{\rm I}$ and $V_{0}^{\rm I}<0$. This negative charge is
completely
screened by positive charge induced in
the region $R<r\leq R_{\rm W}$.

Then we calculate contribution to the thermodynamic potential
(effective energy)
of the Wigner-Seitz cell
per droplet volume. 
We start with the proper electric field energy term;
$\widetilde{\epsilon}_{V}=\widetilde{\epsilon}_{V}^{\,\rm I}
+\widetilde{\epsilon}_{V}^{\,\rm II}$,
\begin{eqnarray}\label{emV}
 \widetilde{\epsilon}_{V}^{\,\rm I}
  &=& \frac{3}{4\pi R^3}\int_{0}^R
	\frac{(\nabla V^{\rm I})^2}{8\pi e^2}4\pi r^2 dr\nonumber\\
  &\simeq& \beta_0 
	\frac{\left(1+\alpha_0 \xi \mbox{th}(\alpha_1 \xi )\right)^2
	\left(-\frac{1}{\xi}\mbox{th}^2 \xi
	+\frac{\xi}{2}-\frac{\xi}{2}\mbox{th}^2 \xi 
	+\frac{1}{2}\mbox{th} \xi\right)}{\xi^3
	\left( \alpha_0\mbox{th}\xi \cdot 
	  \mbox{th}(\alpha_1 \xi ) +1\right)^2},
\end{eqnarray}
being expressed in dimensionless units
\begin{eqnarray}\label{param}
\xi =\kappa^{\rm I}R , \,\,\,\,
\alpha_1 =\frac{\alpha_0 (1-f^{1/3})}{f^{1/3}},
\,\,\,\,f^{1/3}=\frac{R}{R_{\rm W}},\,\,\,\,
\beta_0 = \frac{3\left(U_0^{\rm II}-U_0^{\rm I}
\right)^2 (\kappa^{\rm I})^2}{8\pi e^2},
\end{eqnarray}
where we used eqs.~(\ref{eV}), (\ref{1-sol}), (\ref{const1}).
With the help of  eqs.~(\ref{psi}), (\ref{const2}),
from eq. (\ref{eV}) we find
\begin{eqnarray}\label{ebV}
\widetilde{\epsilon}_{V}^{\,\rm II}
 &=& \frac{3}{4\pi R^3}\int_{R}^{R_{\rm W}}
	\frac{(\nabla V^{\rm II})^2}{8\pi e^2}4\pi r^2 dr \nonumber\\
 &\simeq& \beta_0
\frac{\left(1-\frac{1}{\xi}\mbox{th}
\xi \right)^2
\left(1-\frac{1}{2}\alpha_0 \alpha_1 \xi^2(1-\mbox{th}^2 (\alpha_1 \xi ))
+\frac{1}{2}\alpha_0 \xi\mbox{th} (\alpha_1 \xi )\right)
}{\xi^2
\left( \alpha_0
\mbox{th}\xi \cdot \mbox{th}(\alpha_1 \xi ) +1\right)^2}.
\end{eqnarray}

In order  to
explicitly calculate correlation terms we introduce the quantity $\delta
V^{\rm I}
\equiv V^{\rm I}(r)+\mu_e$ for $r<R$, thus taking
$V^{\rm I}_{\rm r} =-\mu_e$.
Averaging (\ref{om-cor}) over the droplet volume, with the help of
(\ref{1-sol}), (\ref{const1}), we obtain
\begin{eqnarray}\label{emVc}
&&\widetilde{\omega}_{\rm cor}^{\rm I} =
\frac{3}{4\pi R^3}\int_{0}^R
4\pi r^2 dr\omega^{\rm I}_{\rm cor} \simeq
\frac{\beta_0}{2\xi^3 }
\frac{
\left(
1+
\alpha_0 \xi \mbox{th}(\alpha_1 \xi )
\right)^2
}{
\left(  \alpha_0
\mbox{th}\xi \cdot \mbox{th}(\alpha_1 \xi ) +1\right)^2
}\left(\mbox{th}\xi -\frac{\xi}{\mbox{ch}^2 \xi} \right)\nonumber \\
&&
+\frac{2\beta_0}{\xi^3 }\frac{
\left(
1+
\alpha_0 \xi \mbox{th}(\alpha_1 \xi )
\right)
}{\left(  \alpha_0
\mbox{th}\xi \cdot \mbox{th}(\alpha_1 \xi ) +1\right)}(\mbox{th}\xi -\xi )
+\frac{\beta_0}{3 } .
\end{eqnarray}
In the hadron phase introducing
$\delta V^{\rm II}=V^{\rm II}(r)-V^{\rm II}(r=R_{\rm W})$,
where we used $V^{\rm II}_{\rm r} =V^{\rm II}(r=R_{\rm W})\simeq -\mu_e$,
with the help of
eqs.~(\ref{om-cor}), (\ref{psi}),
(\ref{hscrl}), (\ref{const2}),
we obtain
\begin{eqnarray}\label{ebVc}
&&\widetilde{\omega}_{\rm cor}^{\rm II} =\frac{3}{4\pi R^3}\int_{R}^{R_{\rm W}} 4\pi r^2 dr
\omega^{\rm II}_{\rm cor} \simeq  \\
&&\frac{\beta_0 \alpha_0}{6}
\frac{
\left( 1-\frac{1}{\xi}\mbox{th} \xi
\right)^2 \left[ \frac{1}{\xi}\mbox{th} (\alpha_1 \xi )+
3\alpha_1 (1-\mbox{th}^2 (\alpha_1 \xi ))
-\frac{4}{\xi}\mbox{th}
(\alpha_1 \xi )/
\mbox{ch} (\alpha_1 \xi )\right]
}{
\left( \alpha_0
\mbox{th}\xi \cdot \mbox{th}(\alpha_1 \xi ) +1\right)^2
} .\nonumber
\end{eqnarray}
One can see that $\widetilde{\omega}^{\rm II}_{\rm cor} \rightarrow 0$, if
$\alpha_1 \rightarrow 0$, and also in the case $\alpha_0 \rightarrow 0$.
We could also use other values for $V^{\rm I}_{\rm r}$ and
$V^{\rm II}_{\rm r}$, e.g. we have checked that using general eq. (\ref{om-cor0}) with $V^{\rm I}_{\rm r}= V^{\rm I}(0)$
and $V^{\rm II}_{\rm r}= V^{\rm II}( R_{\rm W})$ leads to the very
same result.

In our dimensionless units the total quark plus hadron
surface contribution to the energy per droplet volume renders
\begin{eqnarray}\label{esV}
\widetilde{\epsilon}_{S}/\beta_0 =\beta_1 /\xi ,\,\,\,\, \beta_1
=3\kappa^{\rm I}\sigma /\beta_0 ,
\end{eqnarray}
see  (\ref{param}), and we used that $\epsilon_{S} = 3\sigma/R$.
Coefficients $\beta_0$, $\beta_1$  are evaluated
with the help of  eqs.~(\ref{qscrl}), (\ref{param}) and (\ref{esV}).
For the above used quantities $\mu_{e}\simeq 170~$ MeV,
$\mu_{n} \simeq
1020~$ MeV, $\alpha_c \simeq 0.4$ and $m_s \simeq 120\div 150$~MeV
we estimate
$\beta_0 \simeq 1.6 m_{\pi}^4$. Thus, with the value $\sigma
\simeq 1.3
m_{\pi}^3$ we obtain $\beta_1 \simeq 0.7 $, whereas with $\sigma
\simeq 10~$ MeV/fm$^2 \simeq 0.14m_{\pi}^3$ we would get $\beta_1 \simeq 0.08$.

Coulomb solution for the case of a tiny quark fraction volume is obtained,
if we first put $\alpha_1
\rightarrow \infty$, and then expand the terms
$\widetilde{\epsilon}_{V}^{\,\rm I}+\widetilde{\epsilon}_{V}^{\,\rm II}
+\widetilde{\epsilon}_{S}$ in $\xi \ll 1$.
Thus, we
recover the Coulomb
plus surface
energy per droplet volume
\begin{eqnarray}\label{CS}
\widetilde{\epsilon}_{{\rm C},S}=\widetilde{\epsilon}_{\rm C}+\widetilde{\epsilon}_{S} =
\beta_0 \left( \frac{1}{45}\xi^2 +\frac{1}{9}\xi^2+
\frac{\beta_1}{\xi} \right) ,
\end{eqnarray}
where  partial contributions correspond to the terms
$\widetilde{\epsilon}_{V}^{\,\rm I}$, $\widetilde{\epsilon}_{V}^{\,\rm II}$ and
$\widetilde{\epsilon}_{S}$.
Both the correlation terms
$\widetilde{\omega}_{\rm cor}^{\rm I}\propto \xi^4$
and $\widetilde{\omega}_{\rm cor}^{\rm II}\propto \xi^3$
can be dropped in the Coulomb limit,
for droplets of a tiny size $\xi \ll 1$.

Function $\widetilde{\epsilon}_{{\rm C},S}$ has the minimum at $\xi =\xi_m
=(15\beta_1 /4)^{1/3}$,
corresponding to the optimal size of the
unscreened droplet.
Coulomb solution
is reproduced only for $\xi_m \ll 1$, whereas with above estimate $\beta_1
\gsim 0.1$ we always get  $\xi_m \sim 1$.
On the other hand, for $\beta_1
\ll 0.1$ we would obtain $\kappa^{\rm I}\gsim m_{\pi}$,
corresponding to unrealistically small
droplet size $R< 1/m_{\pi}$.  {\em{Thus, we conclude that pure
Coulomb solution is never  realized within mixed phase.}}

In the limit $\alpha_1 \xi \gg 1$, $\xi \gg 1$,
corresponding to the  single large size drop,
from (\ref{emV}), (\ref{ebV}), (\ref{emVc}) and (\ref{ebVc})
we find that
all the terms contribute to the surface energy
density ($\propto \xi^{-1}$) and, therefore,
the electric field effects can be treated
with the help of an effective surface tension. The full surface
tension $\sigma_{\rm tot}^{\rm spher}$ then renders
\begin{eqnarray}\label{suftenV}
 \sigma_{\rm tot}^{\rm spher}=\sigma +\sigma_{ V}
 =\sigma -
 \lambda_{D}^{\rm I}\frac{\beta_0 \alpha_0
 [\alpha_0 +4/3 ] }{3(1+\alpha_0 )^2} .
\end{eqnarray}
The first $\sigma$ term is the contribution of the strong
interaction, and the second negative term is the contribution of
the electric field effects,
which depends largely on the values of parameters.
For  $\mu_{e}=170$~ MeV,
$\mu_{n}=1020$~MeV, $m_s =150$~MeV, $\alpha_c =0.4$,
we estimate the contribution to the surface tension
from the electric effects
as $\sigma_V \simeq - 70$~MeV$/$fm$^2$.

In Figure
we demonstrate dependence of the contribution
of inhomogeneous charge distributions
to the total thermodynamic potential
per droplet volume (solid curves),
$\delta \widetilde{\omega}_{\rm tot}/\beta_0=(\widetilde{\epsilon}_V +
\widetilde{\omega}_{\rm cor}^{\rm I}+
\widetilde{\omega}_{\rm cor}^{\rm II}+\widetilde{\epsilon}_S )/\beta_0$,
given by the sum of partial contributions
(\ref{emV}), (\ref{ebV}),  (\ref{emVc}),  (\ref{ebVc}) and (\ref{esV}),
for the case of a single droplet,
as function of the droplet size $\xi$
for three values of $\alpha_0$ at fixed value $\beta_1$ (each panel).
The curves labeled by ``C'' demonstrate the Coulomb solution
$\widetilde{\epsilon}_{{\rm C},S} /\beta_0$,
determined by eq.(\ref{CS}).
Each Coulomb curve has a pronounced minimum at
$\xi =\xi_{\rm C} \propto \beta_1^{1/3}$.
For $\xi >\xi_{\rm C}$  the Coulomb curve shows quadratic growth
deviating drastically for $\xi >1$ from the solid curve.
For $\beta_1 \lsim 0.01$,
minimum points of the Coulomb curves $\xi_{\rm C}$
deviate little from the minima of the solid ones.
Only for
such small values of $\beta_1$ and $\xi_m$ we recover
the Coulomb limit!
However, one may obtain such small values of $\beta_1$ only
for tiny values of surface tension and very large values of
neutron chemical potential.
With increase of the latter,
the Debye parameter $\kappa^{\rm I}$ is also increased
and the droplet radius $R =\xi_m /\kappa^{\rm I}$ is proved to be
essentially smaller than $1/m_{\pi}$.
For larger values of $\beta_1$
deviation between  the minima of
total and Coulomb solutions is proved to be pronounced.
All the solid curves converge to $1/3$.
Large $\xi$ asymptotic of solid curves is $\propto 1/\xi$ being
interpreted as the surface energy term, characterizing by a
significantly smaller value of surface tension (\ref{suftenV})
than that determined only by strong interaction.
We see that for
$\sigma \geq \mid \sigma_V \mid$ that corresponds to
$\beta_{1}>\beta_{1{\rm c}}$ ( according to our  Figure
$\beta_{1{\rm c}}\simeq 0.5$ ) the structured mixed phase is
proved to be prohibited, since {\em{ necessary condition}} of its
existence (presence of minimum in the droplet size) is not
satisfied, whereas with the Coulomb solution {\em{necessary
condition}} is always performed. Large size drops are  realized
within the mixed phase if $\sigma < \mid \sigma_V \mid$,
$\mid \sigma + \sigma_V \mid \ll \mid \sigma_V \mid$ and within Maxwell
construction if $\sigma \geq \mid \sigma_V \mid$.
The ``e.m.'' curve shown in each panel
demonstrates the contribution
$\widetilde{\epsilon}_{em}/\beta_0\equiv(\widetilde{\epsilon}_V +
\widetilde{\epsilon}_S)/\beta_0$,
ignoring correlation terms for $\alpha_0 = 1$.
This quantity $\widetilde{\epsilon}_{em}$ is 
the counterpart of the Coulomb solution $\widetilde{\epsilon}_{C,S}$ when 
the charge screening effect is taken into account.
We see that
the minimum at the "e.m." curve disappears already at $\beta_1 > 0.03$.
Difference between the solid and "e.m" curves shows
the important contribution of correlation energy in the hadron-quark 
structured mixed phase.

Dependence of the curves on the ratio
of the screening lengths $\alpha_0$ is also rather pronounced,
whereas it was completely absent for the Coulomb solution.
Our calculations  also show that dependences  of
$\delta \widetilde{\omega}_{\rm tot}/\beta_0$
on the volume fraction $f$
are very weak in the whole range of available values.

Summarizing, in discussion of possibility of presence of structured
mixed phase at first
order phase transitions in multi-component systems of charged particles
we consistently incorporated effects of
the charge screening.
As an example,
our formalism was applied to the hadron-quark structured mixed phase.
We demonstrated that the charge screening effect should 
greatly modify
the description of the mixed phase,
changing its parameters and affecting the possibility of existence.
In absence of the mixed phase
our charged distributions describe
the boundary layer between two separated phases existing
within the double-tangent
(Maxwell) construction.
Consideration of non-spherical droplets (rods and
slabs) does not change our conclusions.
Further discussions are given in 
another paper\cite{vos}.

{\bf{Acknowledgements.}}
Research of D.N.V. at Yukawa Insitute for
Theoretical Physics was
sponsored by the COE program of
the Ministry of Education, Science, Sports and Culture of Japan.  D.N.V.
acknowledges
this support and kind hospitality of the Yukawa Insitute for
Theoretical Physics. He thanks
Department of Physics at Kyoto University
for the warm hospitality and also acknowledges
hospitality and support of GSI Darmstadt.
The present research of T.T. is
partially supported by the REIMEI Research Resources of Japan Atomic Energy
Research Institute, and by the Japanese Grant-in-Aid for Scientific
Research Fund of the Ministry of Education, Culture, Sports, Science and
Technology (11640272, 13640282). We also acknowledge N. Glendenning  and D. Blaschke for
valuable remarks.

\begin{figure}[htb]
 \vspace{2mm}
 \epsfsize=0.99\textwidth
 \epsffile{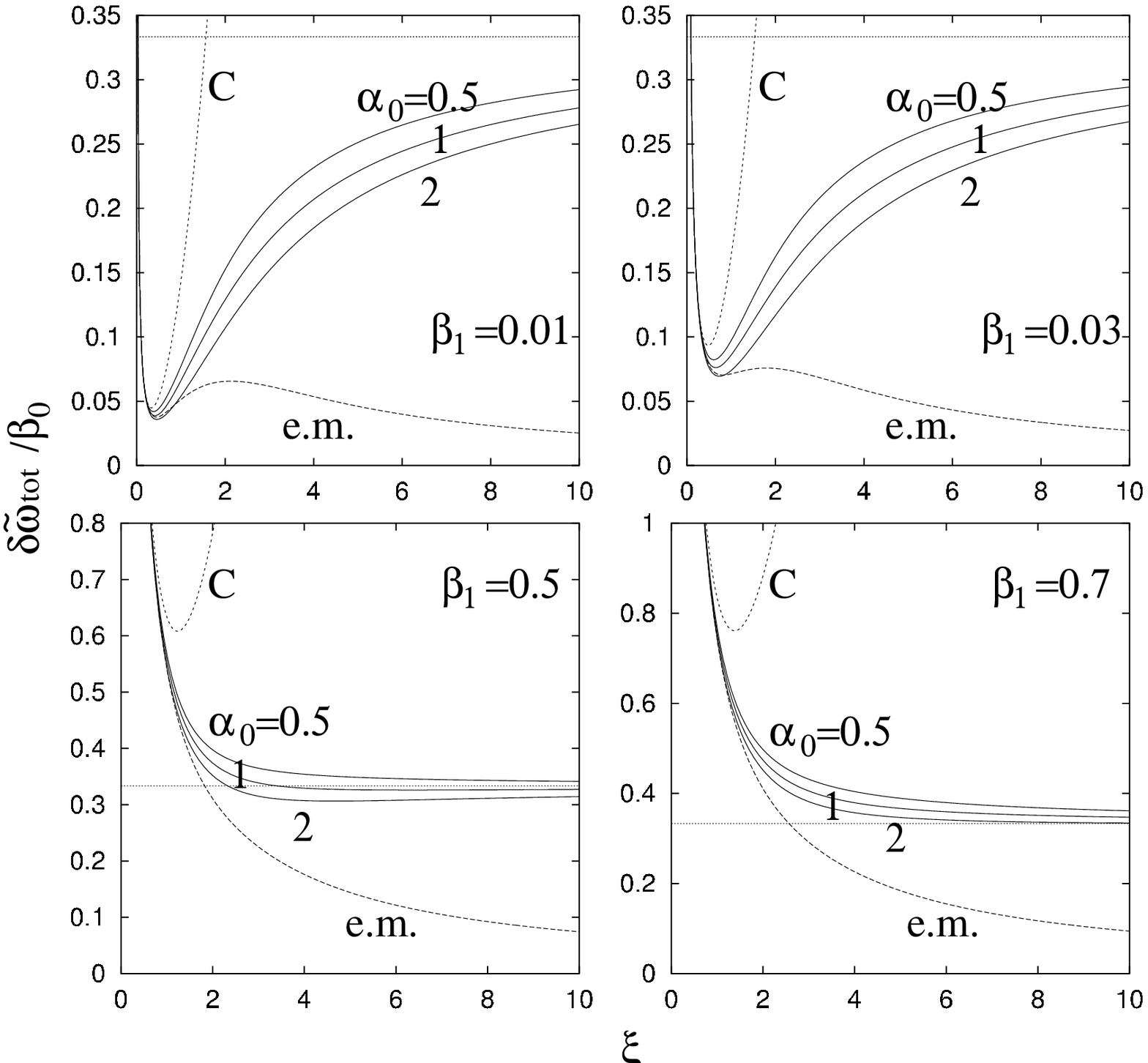}
Figure: Contribution to the effective energy
due to inhomogeneous charge distributions
per droplet volume
in case of a single droplet versus scaled droplet radius.
\end{figure}

\end{document}